\documentclass[final,5p,times,twocolumn]{elsarticle}

\usepackage{amssymb}

\journal{Nucl. Instr. Meth. A}

\begin{document}

\begin{frontmatter}



\title{Recent developments in data reconstruction for aerogel RICH at Belle II}




\author[ULJ,IJS]{L.~\v{S}antelj}       
\author[KEK,SOK]{I.~Adachi}         
\author[IJS]{K.~Adamczyk}               
\author[LAL]{L.~Burmistrov}         
\author[ULJ,IJS]{R.~Dolenec}               
\author[TOK]{K.~Furui}                
\author[NAG]{T.~Iijima}                
\author[TMU]{S.~Iwaki}            
\author[TMC,TMU]{S.~Iwata}        
\author[AANL]{G.~Ghevondyan}           
\author[INA]{R.~Giordano}              
\author[TMU]{H.~Kakuno}            
\author[AANL]{G.~Karyan}           
\author[CHU]{H.~Kawai}             
\author[KIU]{T.~Kawasaki}          
\author[SOK]{H.~Kindo}             
\author[KEK]{T.~Kohriki}             
\author[KIU,TMU]{T.~Konno}             
\author[UMB,IJS]{S.~Korpar}        
\author[ULJ,IJS]{P.~Kri\v{z}an}    
\author[TMU]{T.~Kumita}            
\author[IPMU]{Y.~Lai}               
\author[IJS]{A.~Lozar}               
\author[KEK,SOK,NAG]{K.~Matsuoka}       
\author[TMU]{K.~Motohashi}            
\author[AANL]{G.~Nazaryan}      
\author[KEK,SOK]{S.~Nishida}       
\author[KEK]{M.~Nishimura}       
\author[NIU]{K.~Ogawa}             
\author[TOU]{S.~Ogawa}             
\author[TMU]{T.~Park}         
\author[IJS]{R.~Pestotnik}         
\author[NIU]{Y.~Seino}              
\author[IJS]{A.~Seljak}               
\author[IJS]{L.~Senekovi\v{c}}               
\author[KEK]{M.~Shoji}               
\author[TMU]{T.~Sumiyoshi}         
\author[CHU]{M.~Tabata}            
\author[TMU]{M.~Tsurufuji}         
\author[NIU]{K.~Uno}              
\author[KEK]{E.~Waheed}             
\author[TMU]{K.~Watanabe}      
\author[TMU]{M.~Yonenaga}          
\author[NIU]{Y.~Yusa}              

\address[KEK]{High Energy Accelerator Research Organization (KEK), Tsukuba, Japan}
\address[SOK]{SOKENDAI (The Graduate University of Advanced Studies), Tsukuba, Japan}
\address[LAL]{Laboratoire de l'acc\'e\'lerateur Lin\'eaire (LAL), Orsay, France}
\address[TOK]{University of Tokyo, Tokyo, Japan}
\address[NAG]{Nagoya University, Nagoya, Japan}
\address[TMC]{Tokyo Metropolitan College of Industrial Technoligy, Tokyo, Japan}
\address[TMU]{Tokyo Metropolitan University, Hachioji, Japan}
\address[INA]{Universit\`a di Napoli "Federico II" and Istituto Nazionale di Fisica Nucleare, Napoli, Italy}
\address[AANL]{Alikhanyan National Science Laboratory, Yerevan, Armenia}
\address[CHU]{Chiba University, Chiba, Japan}
\address[KIU]{Kitasato University, Sagamihara, Japan}
\address[UMB]{University of Maribor, Slovenia}
\address[IJS]{Jo\v{z}ef Stefan Institute, Ljubljana, Slovenia}
\address[ULJ]{University of Ljubljana, Slovenia}
\address[IPMU]{KavliI Institute for the Physics and Mathematics of the Universe, Japan}
\address[NIU]{Niigata University, Niigata, Japan}
\address[TOU]{Toho University, Funabashi, Japan}

\begin{abstract}
  In the forward end-cap of the Belle II spectrometer, particle identification is provided by a proximity focusing RICH detector with an aerogel radiator (ARICH). The ARICH's primary function is to effectively distinguish between pions and kaons in the momentum range of $0.5~\mathrm{GeV}/c$ to about $4~\mathrm{GeV}/c$, as well as to contribute to identification of low-momentum leptons. Since its operation began, Belle II has collected over $420~\mathrm{fb}^{-1}$ of data. Based on this large data sample, studies of several effects that impact the performance of the ARICH detector were carried out. In this paper, we present a comparison of the observed Cherenkov ring image and detector particle identification performance in the measured data and detector simulation. Furthermore, we highlight recent efforts aimed at enhancing the ARICH's performance by taking into account the effects of particle decay in flight and scattering in materials before the detector, as well as by refining the probability density function used for particle identification likelihood evaluation.

\end{abstract}


\end{frontmatter}


\section{Introduction}
\label{intro}
\begin{figure}[h]
  \includegraphics[width=0.34\textwidth]{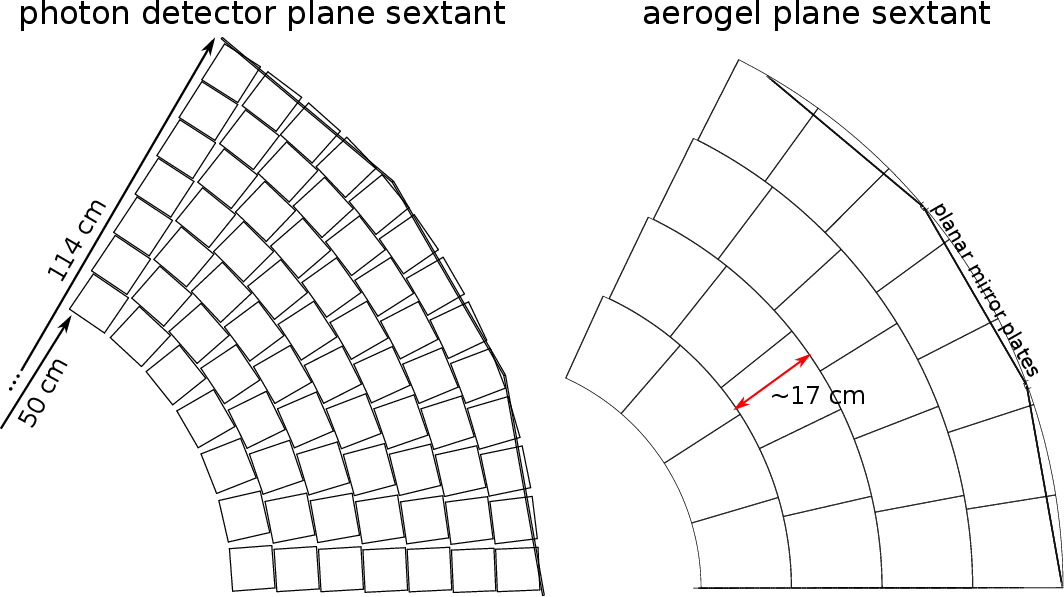}
  \includegraphics[width=0.13\textwidth]{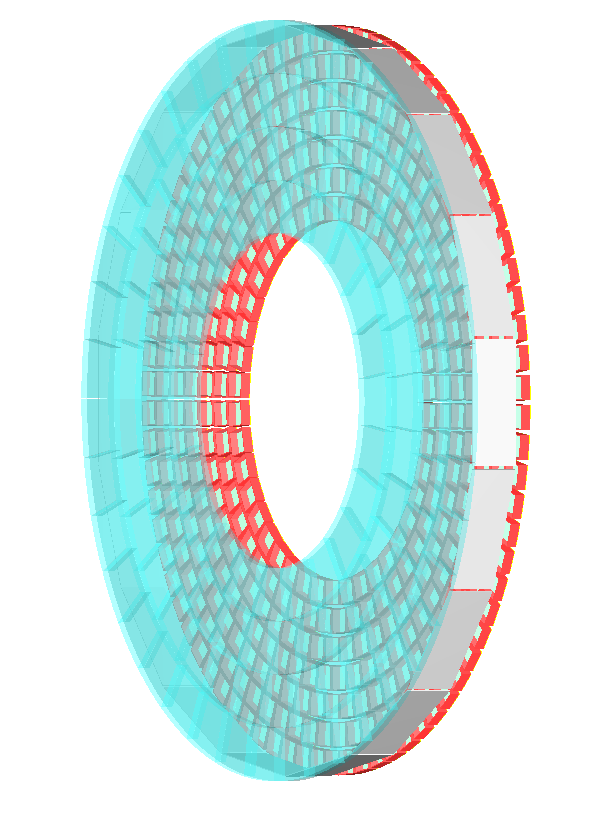}
  \caption{Left: Sketch of distributions of photon detector modules and aerogel tiles in one sextant of the detector. Right: Image of the ARICH detector as exported from the Geant4 simulation.}
  \label{fig:geo}
\end{figure}

At the Belle II experiment \cite{b2}, a proximity focusing RICH detector with an aerogel radiator (ARICH) \cite{arich} is used to provide particle identification in the forward end-cap of the spectrometer. The detector's primary purpose is to effectively separate pions and kaons in the full kinematic range of the experiment, from $0.5-4.0~ \mathrm{GeV}/c$. The detector is located approximately $170~\mathrm{cm}$ from the interaction point and consists of a radiator plane, an expansion volume of approximately $17~\mathrm{cm}$, and a photon detector plane. The radiator plane is covered with two layers of aerogel tiles, each $2~\mathrm{cm}$ thick and wedge-shaped with dimensions of around $17~\mathrm{cm}$. The average refractive indices and transmission lengths of the tiles in the first and second layers are 1.045 and $45~\mathrm{mm}$, and 1.055 and $35~\mathrm{mm}$, respectively. The photon detector plane comprises 420 Hybrid Avalanche Photo-Detectors (HAPDs) \cite{hapd}, arranged in 7 concentric rings. Each HAPD has a sensitive surface of about $6\times 6~\mathrm{cm}$ and is pixelated into $12\times 12$ pixels with a pitch of $0.5~\mathrm{cm}$. The average quantum efficiency of the photocathode is approximately 30\% at $400~\mathrm{nm}$ and the overall coverage of the photon detector plane is about 60\%. The detector geometry is sketched on Figure \ref{fig:geo}. Belle II began collecting data in 2019 and has so far collected $420~\mathrm{fb}^{-1}$ of data, mainly at the energy of $10.58~\mathrm{GeV}$, corresponding to the energy of the $\Upsilon(4S)$ resonance.

In this letter, we present a comparison of the detector response and its particle identification capabilities as observed in measured data with that from a detailed detector simulation. Additionally, we report on few selected recent developments in data reconstruction methods that we expect will improve the detector performance in the near future.

\section{Measured data vs. detector simulation}

\begin{figure}[ht!]
  \includegraphics[width=0.3\textwidth]{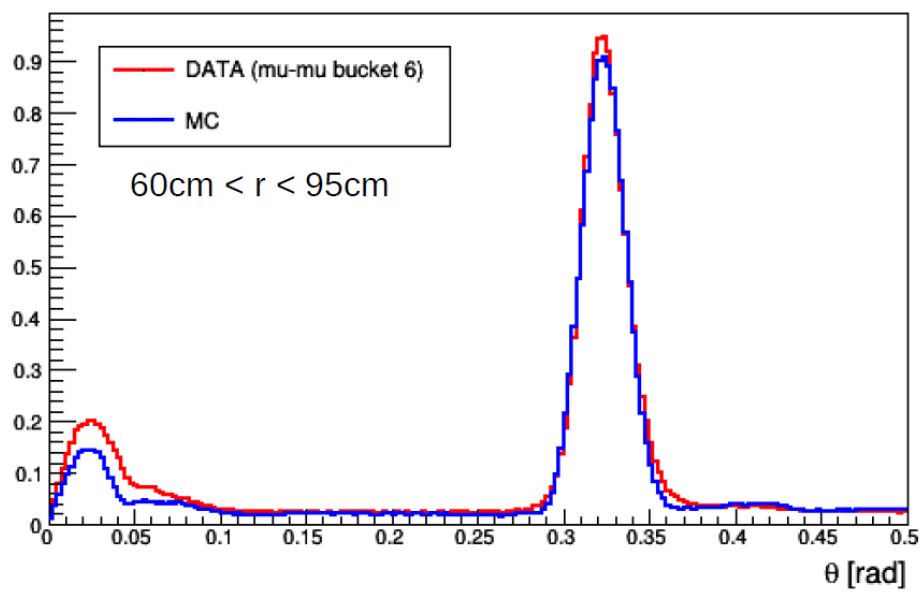}
  \includegraphics[width=0.16\textwidth]{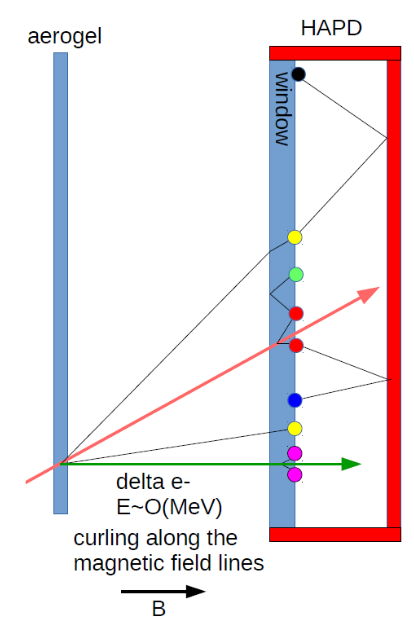}
  \includegraphics[width=0.5\textwidth]{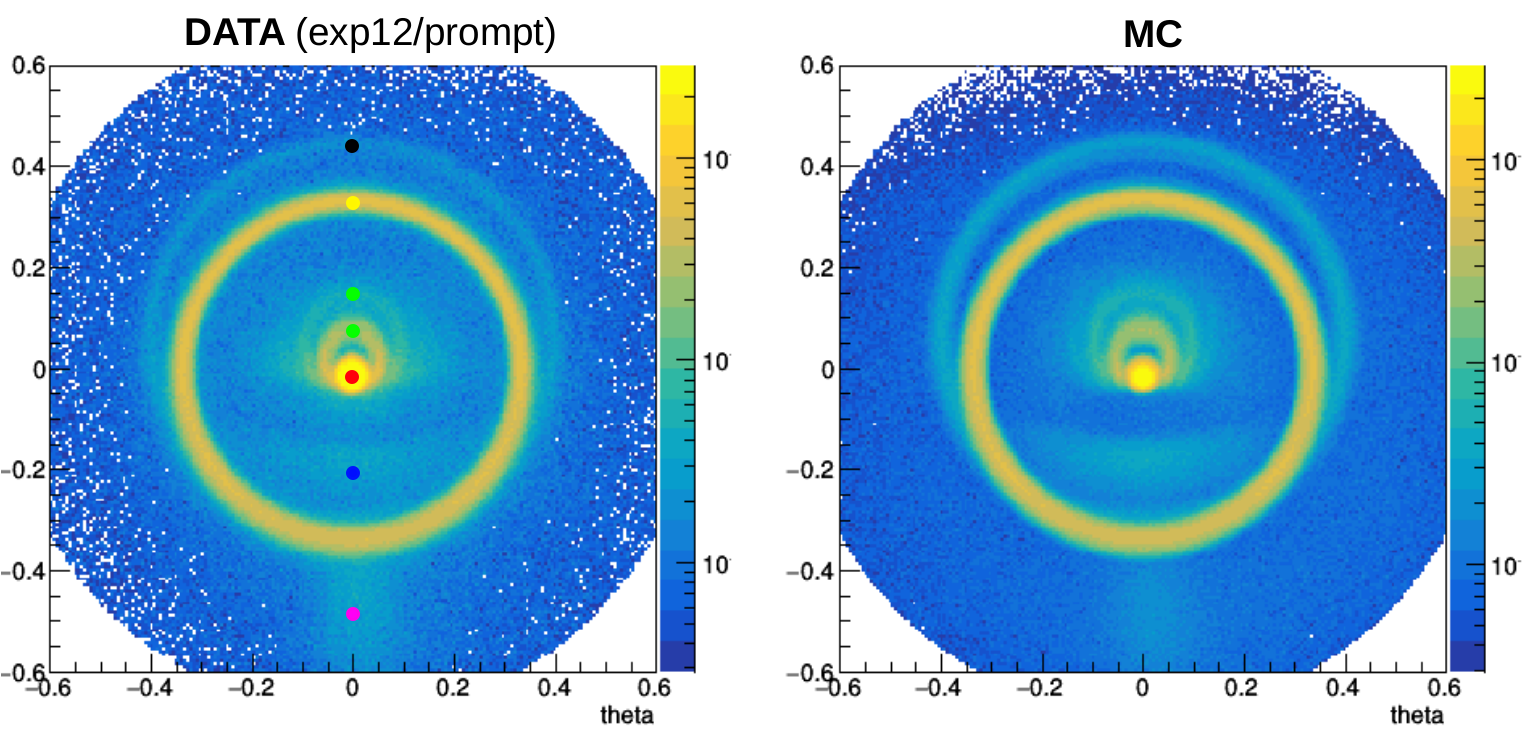}
  \caption{Accumulated Cherenkov ring image for high energy muons (from $e^+e^-\to \mu^+\mu^-$ events) as observed in the measured data and detector simulation (MC). The origin of several features of the ring image is explained by the sketch on the upper right and the correspondingly colored dots on the ring image from the measured data.}
  \label{fig:ring}
\end{figure}

We compare the measured detector response upon a passage of high energy muon from $e^+e^-\to \mu^+\mu^-$ events with the corresponding response from a detailed Geant4 simulation of the full Belle II detector. Muons from these events have momenta around $6.5~\mathrm{GeV}/c$ and thus produce fully saturated Cherenkov rings in aerogel. Figure \ref{fig:ring} shows the accumulated distribution of reconstructed Cherenkov angle normalized to a single track. By fitting a Gaussian function and flat background to the observed distributions we determine the number of photons per track and single photon Cherenkov angle resolution to be 11.4 and $12.7~\mathrm{mrad}$ for the measured data, and 11.3 and $12.8~\mathrm{mrad}$ for the simulated data. Notably, a slight discrepancy is observed in the peak at the smallest Cherenkov angles, which originates from the mismodeling of the Cherenkov photons produced in the quartz window of HAPD. This will be further tuned in the next productions of the simulated data to better reproduce the measured data. In addition, on Figure \ref{fig:ring}, we show the full ring image in the Cherenkov space. Besides the main Cherenkov ring, we identified several subtle features of the image and tried to understand their origins. Several of these features originate from possible Cherenkov photon reflections in the HAPD either from internal reflections in its quartz window or reflections of non-converted photons from the APD silicon chip back to the photocathode. Interestingly, one can also see the contribution from $\delta$-electrons produced by charged particles in the aerogel, which propagate along the magnetic field lines to the photon-detector plane and produce Cherenkov photons in the HADP window. The upper right sketch of Figure \ref{fig:ring} schematically shows the explanation of each of the identified features. All of these features are nicely reproduced in the simulated data, where we confirm that our understandings are correct.

\begin{figure}[ht!]
  \includegraphics[width=0.5\textwidth]{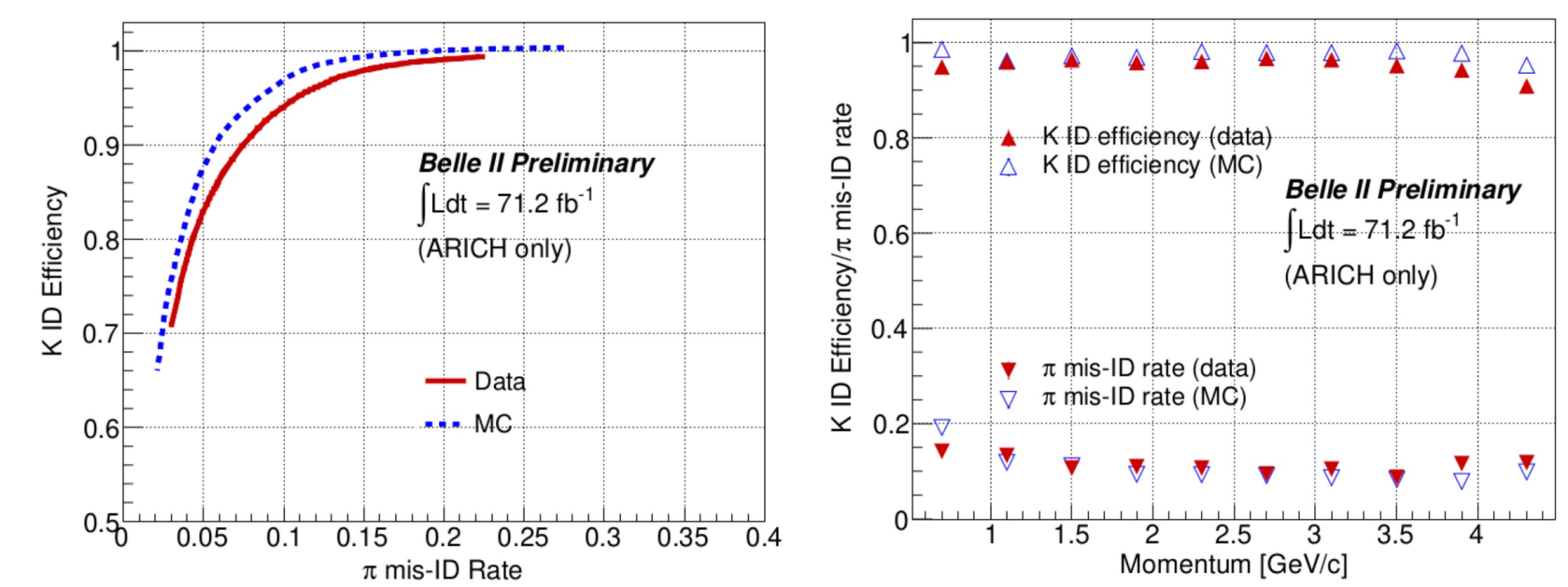}
  \caption{ARICH kaon/pion identification performance as observed in the measured data and in the simulation. The left plot includes all tracks from $D^0\to K^-\pi^+$ decays with momenta from $0.5 - 3.5~\mathrm{GeV}/c$.}
\label{fig:perf}
\end{figure}

We use pion and kaon tracks from the $D^{*+}\to D^0\pi^+_{slow}$ decay followed by $D^0 \to K^-\pi^+$ to evaluate the detector performance. In this case, pions and kaons can be discriminated based on the correlation between their charge and the charge of the slow pion $\pi_{slow}$. The kaon identification efficiency versus pion misidentification probability curve is shown in Figure \ref{fig:perf}. The dependency of kaon/pion discrimination on the track momenta is shown on the right plot of the figure\footnote{here kaon/pion likelihood difference is required to be larger than zero.}. We observe slightly worse performance in the measured data compared to the simulation. We attribute this to the following reasons: 

\begin{itemize}
\item Imperfect description of gaps between neighboring aerogel tiles in the detector simulation and their misalignment. Cherenkov photons are lost on the edges of aerogel tiles, due to absorption or scattering. This loss affects a relatively large fraction of tracks and is taken into account in the construction of the likelihood probability density function (PDF). However, if the gap between two neighboring tiles is misaligned from its nominal position or is of a different width than assumed, the photon loss is not correctly assigned, resulting in increased misidentification probability. We confirm this by demonstrating that the difference in measured and simulated performance is significantly reduced when tracks that hit the aerogel plane near such a gap are excluded from analysis. Currently, an algorithm for individual tile alignment is under development.

\item Possible underestimate of the amount of material in the ARICH in the detector simulation. We observe that the data/simulation performance difference is further reduced when only tracks for which we confirm that they are not scattered in the material (by checking the response of detectors placed behind the ARICH) are taken into account. This suggests that the effect of material scattering is larger in the measured data than in the simulation.

\item Imperfect detector alignments. At high momenta (around 4 GeV/c), PID performance starts to depend on the quality of detector global alignment and alignments of planar mirrors installed on the outer edge of the detector. With larger data sets now available, we aim to improve these alignments which will reduce their impact on performance.

\end{itemize}

The first of the listed is observed to contributute to the data/simulation performance difference  most significantly. 

\section{Recent improvements in the data reconstruction agorithms}

Several efforts to enhance the performance of the ARICH are currently in progress. This section provides more information on two of these efforts: handling of decayed and scattered particles and improvements to the construction of the PID likelihood function.

\subsection{Treatment of decayed/scattered particle}

\begin{figure}[h!]
  \includegraphics[width=0.45\textwidth]{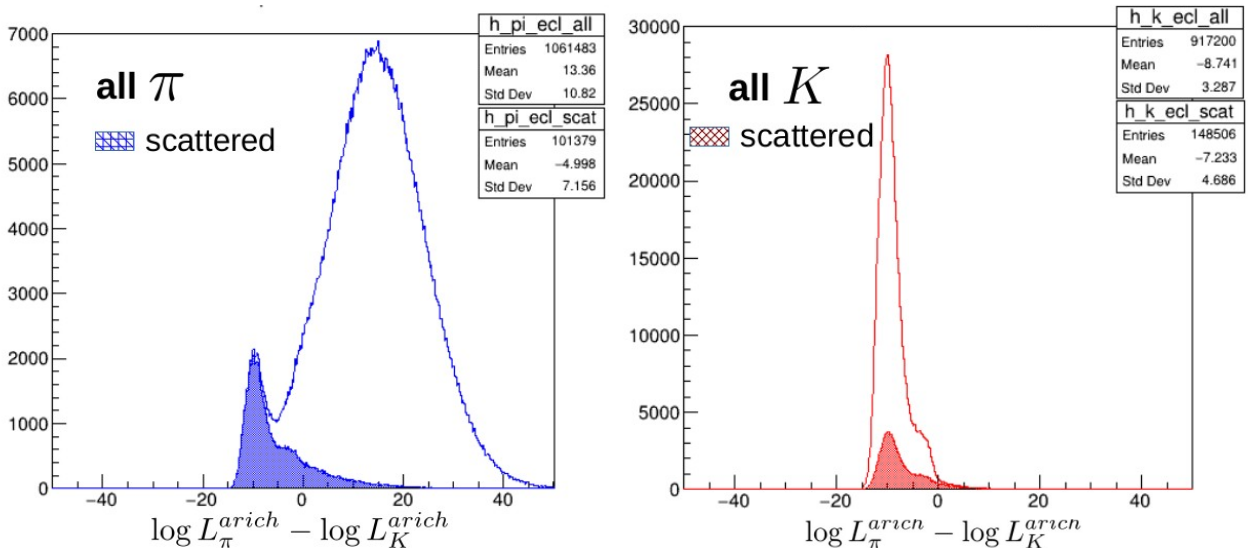}
  \includegraphics[width=0.45\textwidth]{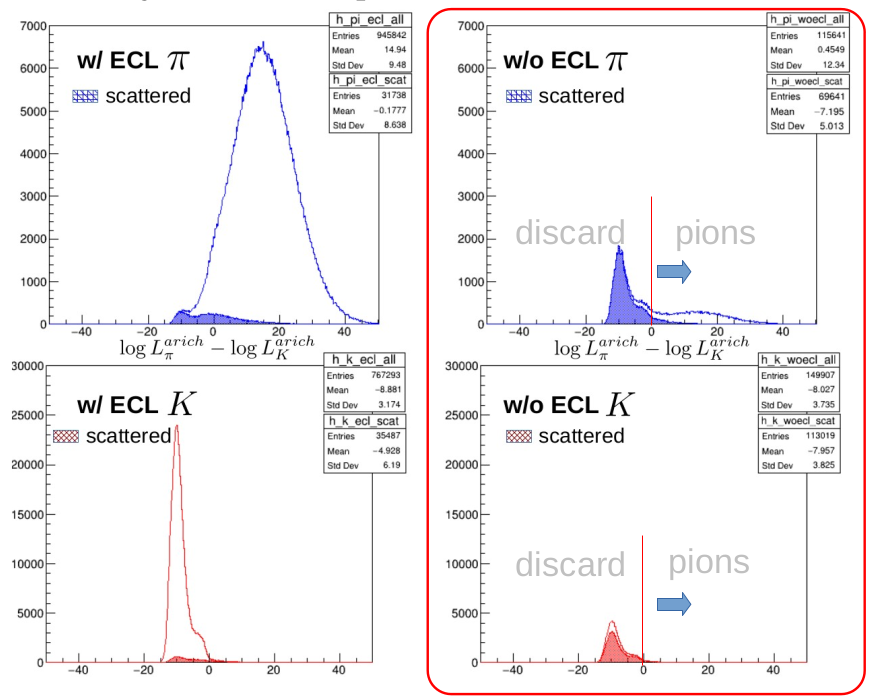}
  \caption{Upper: likelihood difference distribution for simulated pions (left) and kaons (right) at $1~\mathrm{GeV}/c$. Filled histograms show contribution of decayed/scattered particles. Middle: likelihood difference distribution for simulated pions with the sample split into subsamples with (left) and without (right) the associated ECL cluster. Bottom: equivalent as middle, but for kaons.}
\label{fig:dec_lkh}
\end{figure}

In our analysis of simulated data, we discovered that a significant proportion of particles misidentified by the ARICH are from tracks that are extrapolated to the ARICH detector, but the particle that produced the track did not actually reach the ARICH. This occurs as a result of particle decay in flight or scattering in the material in front of the ARICH (most notably in the drift chamber aluminum end-plate). From the perspective of the $\mathrm{Belle~II}$ global PID performance, it would be beneficial if the PID likelihood from the ARICH is not assigned to such tracks, as it often leads to incorrect identification. A notable example are decayed/scattered pions at momenta below the kaon threshold (around $1.5~\mathrm{GeV}/c$), which are due to a lack of photons strongly idenfitied as kaons. This can be observed in Figure \ref{fig:dec_lkh}, which shows the likelihood difference distribution for simulated pions and kaons at a momentum of $1~\mathrm{GeV}/c$ (filled histograms show the contribution of tracks for which we check on the generator level that the particle producing it did not actually pass through the ARICH).

\begin{figure}[h!]
  \centering
  \includegraphics[width=0.35\textwidth]{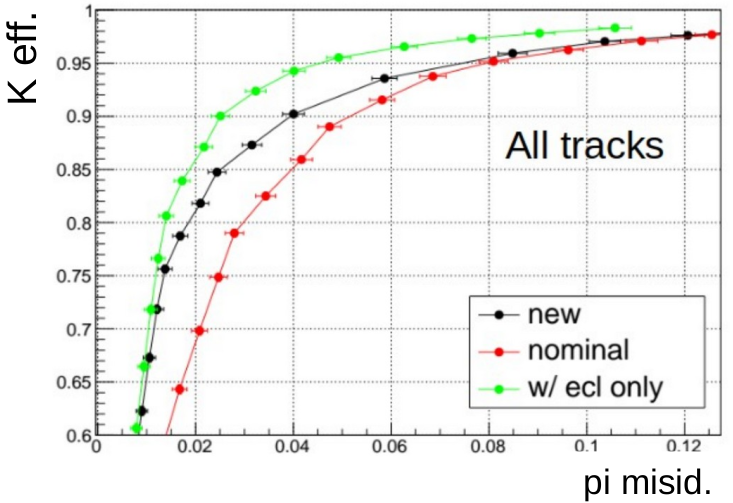}
  \caption{Kaon identification effciency versus pion misidentification probability of the Belle II global PID system for kaons and pions from $D^0\to K^+\pi^-$ decays with extrapolated track in the ARICH. Curves as obtained with the nominal method (i.e. including ARICH likelihoods for all tracks), the new method (as described in the text), and for a subsample of tracks with associated ECL cluster.}
\label{fig:dec_roc}
\end{figure}

To mitigate this effect, we first attempt to identify such decayed/scattered particles based on information from the $\mathrm{Belle~II}$ electromagnetic calorimeter (ECL) which is located immediately behind the ARICH detector. According to the simulated data, about 75\% of these particles do not have an associated ECL cluster. Therefore, we apply a different method for building the global PID likelihood for tracks without the ECL cluster (in total, these represent about 10\% of all tracks). For these tracks, we only include the ARICH likelihoods in the global PID if $(\log{L^{arich}_{\pi}}- \log{L^{arich}_{K})}>0$ and discard the ARICH information otherwise. The motivation for this is obvious from the middle and bottom plots of Figure \ref{fig:dec_lkh}. One can see that among tracks without ECL cluster, basically all tracks identified as pions were true pions, while the ones identified as kaons contain some true kaons but also a large number of misidentified pions. With this simple treatment, we improve the $\mathrm{Belle~II}$ global PID performance as shown in Figure \ref{fig:dec_roc}.

In order to further improve performance, we continue to explore options for improving the identification of decayed/scattered particles by including information from detector systems other than ECL, for example, by studying the track hit patterns in the $\mathrm{Belle~II}$ central drift chamber.

\subsection{Improving PID likelihood PDF}

\begin{figure}[h!]
  \includegraphics[width=0.2\textwidth]{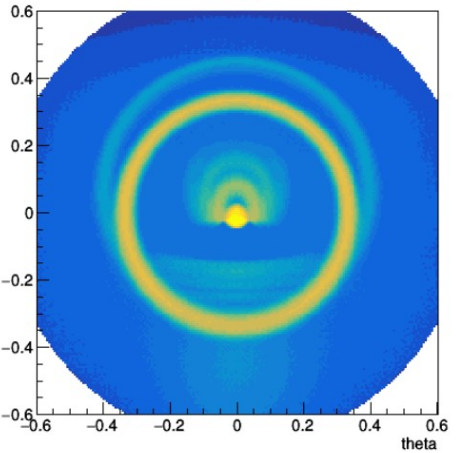}
  \includegraphics[width=0.25\textwidth]{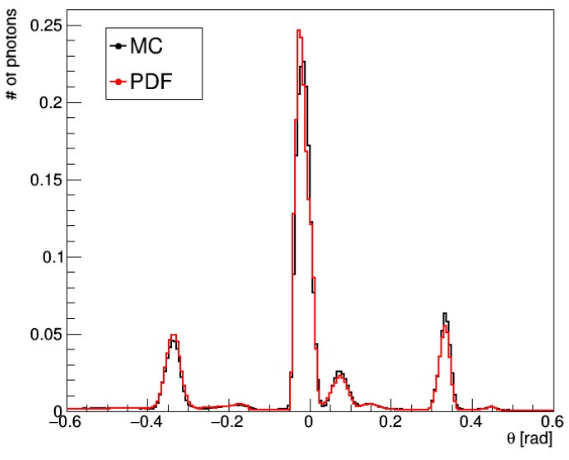}
  \caption{Left: The expected Cherenkov ring image for saturated track as obtained by the improved PDF. Right: Narrow vertical slice of left image (PDF) and comparison with the accumulated ring image observed in the detector simulation (MC).}
\label{fig:pdf}
\end{figure}

Our PDF that describes the expected distribution of photon hits on the photon detector plane for a given track parameters and particle type hypothesis nominaly includes only a description of the main Cherenkov ring and simple background, without any dependence on the ring azimuthal angle. The PDF is constructed on a track-by-track basis by projecting an analytic description of the Cherenkov angle distribution from the Cherenkov space onto the photon detector plane. More details of the procedure are described in \cite{pdf}. However, as seen in Figure \ref{fig:ring}, there are several distinct features that contribute to the ring image. An understanding of the origin of these features enabled us to extend the existing PDF to better describe the observed image. Since the position, shape, and relative intensities of these features depend on the track parameters and detector geometry, they are calculated for each track by performing a simple toy simulation of photon propagation. This simulation includes basic detector geometry, sufficient to obtain the shapes and intensities of contributions from photons generated in the HAPD window, possible reflections, etc. The obtained PDF (accumulated over multiple tracks) is shown in Figure \ref{fig:pdf}, where one can see good agreement with the distribution in the simulated data. It is important to note that the only free parameters of this PDF are the relative intensities of each of the features with respect to the main ring\footnote{These intensities depend on the details of the optical properties of photo-cathode and reflectivity of APD surface}. These parameters are tuned to reproduce the observed image.

\begin{figure}[h!]
  \centering 
  \includegraphics[width=0.3\textwidth]{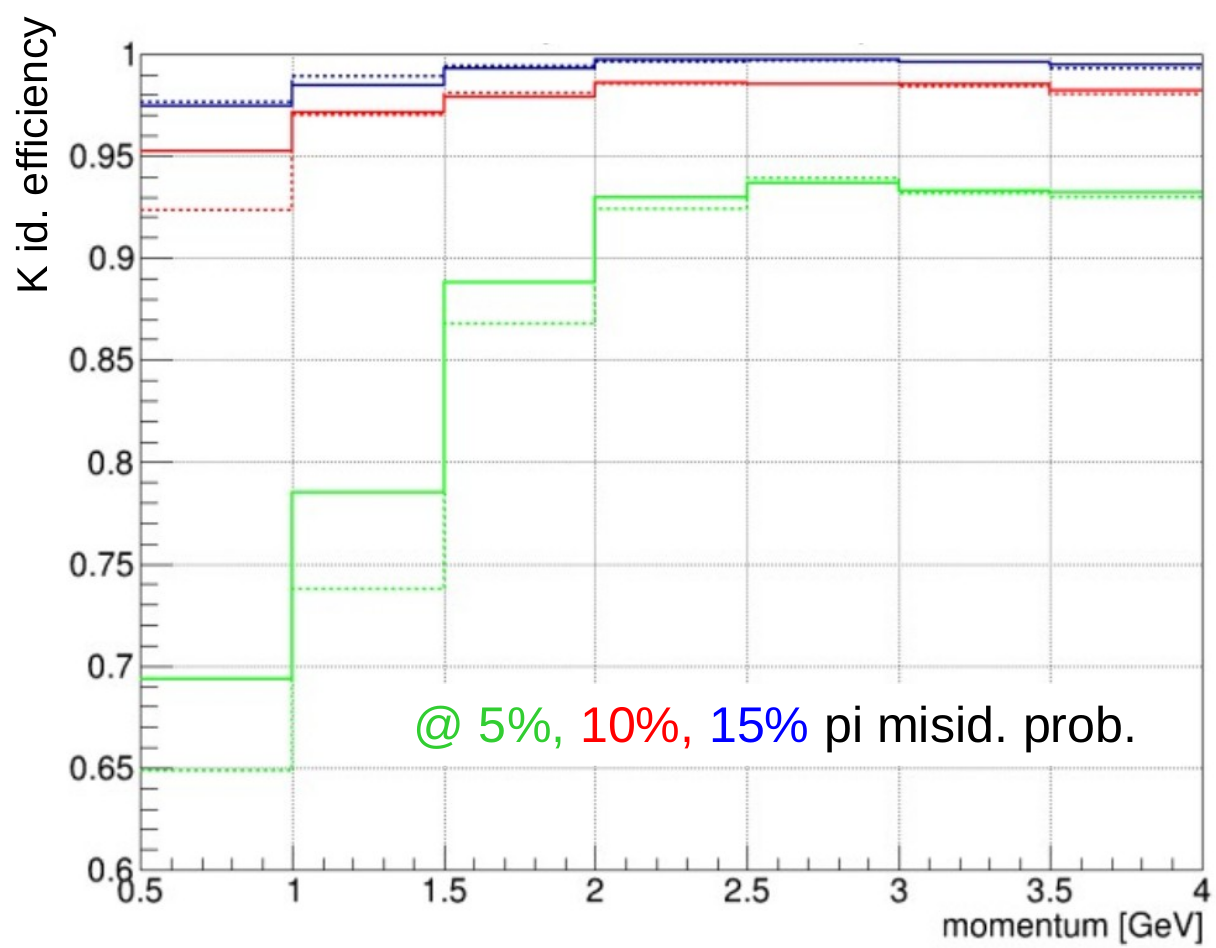}
  \caption{Comparison of kaon identification efficiency at three different pion misidentification probabilities obtained with our nominal (dashed lines) and improved PDF (solid lines).}
\label{fig:pdf_perf}
\end{figure}

In Figure \ref{fig:pdf_perf}, we compare the PID performance obtained using the original and improved PDF in the simulated data. For the low momentum tracks notable improvement in the PID performance is observed, particularly when considering low pion misidentification probabilities. The tuning of the new PDF parameters to the measured data and performance improvement studies are ongoing.

\section{Conclusions and outlook}

The ARICH detector at the Belle II spectrometer is operating stabily and providing particle identification capabilities close to expected from the detector simulations. Neverthelss, as presented in this text, several efforts are ongoing in order to further improve the performance and to reduce the observed discrepancies between the measured and simulated data. Of particular significance, we anticipate notable improvements through the implementation of more advanced techniques for handling particles that decay or scatter before reaching the ARICH detector. Additionally, we are working to refine the description of photon loss on the edges of aerogel tiles in the detector simulation and to provide alignment algorithm for these edges, which will reduce observed data/simulation discrepancies.   
\section{Acknowledgements}

This work was supported by the following funding sources: European Research Council, Horizon 2020 ERC-Advanced Grant No. 884719; Slovenian Research Agency research grants No. J1-9124, J1-4358 and P1-0135 (Slovenia).




\begin{thebibliography}{00}

\bibitem{b2}
  T. Abe {\it et al.} (Belle II Collaboration), (2010), arXiv:1011.0352 [physics.ins-det]

\bibitem{arich}
  T. Iijima, S. Korpar {\it et al.}, Nucl. Instrum. Meth. {\bf A548}, (2005) 383   
\bibitem{hapd}
  S. Korpar {\it et al.}, Nucl. Instrum. Meth. {\bf A766}, (2014) 145

\bibitem{pdf}
  L. \v Santelj {\it et al.}, Nucl. Instrum. Meth. {\bf A876}, (2017) 104
\end{thebibliography}


\end{document}